\documentclass{article}

\usepackage{arxiv}

\usepackage[utf8]{inputenc} 
\usepackage[T1]{fontenc}    
\usepackage{hyperref}       
\usepackage{url}            
\usepackage{booktabs}       
\usepackage{amsfonts}       
\usepackage{nicefrac}       
\usepackage{microtype}      
\usepackage{lipsum}

\usepackage{float}
\usepackage{lscape}
\usepackage{booktabs}
\usepackage{amsmath}
\usepackage{amsfonts}
\usepackage{hyperref}
\usepackage[natbibapa]{apacite}

\usepackage{booktabs}
\usepackage{lscape}

\newcommand{\wxxsir}{eight } 
\newcommand{\Wxxsir}{Eight } 
\newcommand{\rr}{\ensuremath{R_0}} 

\usepackage{systeme}

\usepackage{tikz}
\tikzset{
  int/.style={circle, draw, fill=blue!20, minimum size=3em},
  init/.style={pin distance=1.2cm,pin edge={loop,thin,black}}
}
\usetikzlibrary{arrows,automata}

\title{Exploring the nuances of $\rr$: \Wxxsir estimates and application to 2009 pandemic influenza}

\author{
  Shannon Gallagher\\
  Biostatistics Research Branch, Division of Clinical Research\\
  National Institute of Allergy and Infectious Diseases\\
  Rockville, MD 20852\\
  \texttt{shannon.gallagher@nih.gov} \\
   \And
Andersen Chang\\
  Department of Statistics\
  Rice University\\
  Houston, TX 77005\\
  \texttt{atc7@rice.edu}\\
  \And
  William F. Eddy\\
  Department of Statistics \& Data Science\\
  Carnegie Mellon University\\
  Pittsburgh, PA 15213 \\
}

\begin{document}
\maketitle

\begin{abstract}
For nearly a century, the initial reproduction number ($\rr$) has been used as a one number summary to compare outbreaks of infectious disease, yet there is no `standard' estimator for $\rr$.  Difficulties in estimating $\rr$ arise both from how a disease transmits through a population as well as from differences in statistical estimation method.  We describe \wxxsir methods used to estimate $\rr$ and provide a thorough simulation study of how these estimates change in the presence of different disease parameters.  As motivation, we analyze the 2009 outbreak of the H1N1 pandemic influenza in the USA and compare the results from our \wxxsir methods to a previous study.  We discuss the most important aspects from our results which effect the estimation of $\rr$, which include the population size, time period used, and the initial percent of infectious individuals.  Additionally, we discuss how pre-processing incidence counts may effect estimates of $\rr$.  Finally, we provide guidelines for estimating point estimates and confidence intervals to create reliable, comparable estimates of $\rr$.
\end{abstract}


\section{Introduction}\label{sec:intro}
What has been called ``arguably the most important quantity in the study of epidemics,''  $\rr$ (by convention pronounced ``R-naught''), the initial reproduction number remains an important quantity to estimate to assess the severity of infectious diseases. As defined by \citet{anderson1992}, $\rr$ is the ``the average number of secondary infections produced when one infected individual is introduced into a host population where everyone is susceptible.'' Often understood to be synonymous with the severity of infection, $\rr$ is still used to inform both professionals and lay people alike.  For example, the Wall Street Journal in February 2020 reported an estimate of $\rr$ before the impending outbreak of COVID-19, a novel coronavirus, in the West \citep{wsj2020}.   Because of this reliance upon this estimate, it is more important than ever to produce accurate and reliable estimates of this quantity.

In some ways, $\rr$ summarizes the entire outbreak of a disease. $\rr$ is used to assess whether a disease outbreak will occur and its severity.  Additionally, $\rr$ describes what percentage of the population needs to be vaccinated to avoid such an epidemic, roughly $1-\rr^{-1}$; is used to estimate the final size of the total number of infected individuals; and is related to the probability of observing an outbreak under the same conditions \citep{anderson1992,britton2010}.  Despite a clear definition of $\rr$, epidemiologists have struggled to create a standard  estimator for $\rr$  \citep{hethcote2000}.

 One major issue in estimating $\rr$ is that the quantity is a \textit{property of the model}, meaning that $\rr$ is dependent not only on the usual noise that comes with statistical modeling but also on a variety of assumptions on how researchers assume a disease is transmitted through a population \citep{diekmann2009,brown2016}.  The consequences of this is that despite being nominally the same, $\rr$ may have a different interpretation depending on the model used. As an example, there are two common types of epidemic frameworks to describe the transmission of infectious disease;  SIR and SEIR models.  Here, `S' stands for susceptible, `I' for infectious, `R' for recovered or removed, and `E' for exposed (i.e. infected but not yet infectious).  In both models, $\rr$ is defined in the same manner, namely the infection rate over the recovery rate, $\rr = \frac{\beta}{\gamma}$.  However, the two estimates are not directly comparable to one another.  An analogy is comparing coefficient $\beta_1$ that appears in two linear regression models, where the second model is nested in the first, as shown in Eq. \eqref{eq:lin-reg1}-\eqref{eq:lin-reg2}.
 \begin{align}
 E[Y] &= \beta_0 + \beta_1 X_1 + \beta_2 X_2 \label{eq:lin-reg1}\\
 E[Y] &= \beta_0 + \beta_1 X_1  \label{eq:lin-reg2}.
 \end{align}
 In both models, estimates of $\beta_1$ can indicate whether $X_1$ is a significant variable or not.  However, in Eq. \eqref{eq:lin-reg1}, $\beta_1$ can only be interpreted in the context of $X_2$ whereas the interpretation $\beta_1$ in Eq. \eqref{eq:lin-reg2} has no explicit connection to $X_2$.
 
 A number of papers have been written about the difficulties and nuances involved in estimating $\rr$ (see \cite{hethcote2000,diekmann2009,li2011,driessche2017}.  Our study differs from those previous through a study of different methods to estimate $\rr$, emphasis on the data format, and sensitivity to certain disease parameters. However, even in the same model framework (e.g. SIR models), estimates may differ from one another.  Difference in estimates can result from mathematical versus statistical methods (i.e. solving for $\rr$ versus estimating $\rr$), the number of observations used to estimate $\rr$ (e.g. time steps),  population size, initial SI ratio, and model specification.  These are all issues for producing point estimates, to say nothing of issues related to estimating confidence intervals (CI).  Moreover, there is the important issue of the data itself. Methods to estimate SIR curves typically assume the observed number of Susceptible, Infectious, and Recovered (S(t), I(t), R(t)) are available at a number of time steps, but typically those values are unobserved, and we, instead, observe only newly reported cases in a given time interval.

 In this paper, we focus on highlighting the three following aspects of $\rr$: 1) variety in estimators, 2) sensitivity to certain disease parameters, and 3) how the difference between ideal and actual data can influence our estimates of $\rr$.  We limit our estimates to \wxxsir estimates that are constructed to work within the SIR framework.  Moreover, we consider Bayesian models to be outside the scope of our paper.  Our results are shown in both a simulation study as well as an application to 2009 pandemic influenza.
 
 The rest of this paper is organized as follows.  In Section \ref{sec:methods}, we introduce the deterministic SIR model and stochastic variations of it.  We then describe eight different estimates of $\rr$ in Sections \ref{sec:methods-eg}-\ref{sec:methods-lma}.  Following that in Section \ref{sec:data} we describe the data and nuances thereof of both our simulation data and the 2009 pandemic influenza data.  In Section \ref{sec:results}, we describe the results of the simulation study the 2009 pandemic influenza application.  Finally, in Section \ref{sec:discussion}, we discuss our recommends and conclusions from our analysis.

\section{Methods}\label{sec:methods}

\textbf{The deterministic model}.  In order to discuss our \wxxsir methods to estimate $\rr$ in the SIR model, we first need to define the SIR model. The SIR model introduced by Kermack and McKendrick \citeyearpar{Kermack700} is a compartment model, where individuals move from susceptible, to infectious, and finally recovered states (compartments).  We study the SIR model without vital dynamics (e.g. no birth and death into or out of a population).  We make five essential assumptions: 1) the compartments are discrete and have no overlap, 2) transition of individuals is described by a set of known equations, possibly dependent on an unknown parameter, 3) the populations mix  homogeneously, 4) the number of individuals in each compartment at time $t=0$ is known, and 5) the law of mass action is obeyed. The last of these conditions is a property borrowed from chemistry which says that the mass of the product per unit time is proportional to the mass of the reactants \citep{lotka1920}.  In epidemiology, this means that the proportion of new infections per unit time is proportional to the current  number of susceptible \citep{anderson1992}.

\begin{figure}[H]
\centering
\begin{tikzpicture}[node distance=2cm,auto,>=latex',every node/.append style={align=center}]
    \node [int,  fill = white!70!blue] (a)              {$X$};
    \node [int,  fill = white!70!blue]           (c) [right of=a] {$Y$};
    \node [int,  fill = white!70!blue] (e) [right of=c] {$Z$};
    \path[->, auto=false] (a) edge node {} (c)
                          (c) edge node {} (e) ;
\end{tikzpicture}
\caption{Depiction of a SIR model where $X=S, Y=I,$ and $Z=R$.  One can only be infected once by the disease in this model.  Individuals begin in the X state as susceptible individuals, possibly become infectious and move to the Y state, and finally possibly recover in the Z state.  There are no births or deaths in this SIR model.}\label{fig::sir}
\end{figure}
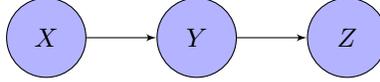

In this SIR model,  (displayed graphically in Figure \ref{fig::sir}), the total number of individuals ($N$) is constant. Adaptations of the SIR model can include birth and death rates, which may correspondingly change the derivation of $\rr$ (for further discussion. Recall, $\beta$ is the average infection rate and $\gamma$ is the average recovery rate.  We assume $\beta$ and $\gamma$ are positive. The movement of individuals from one compartment to another is represented through the ordinary differential equations below.   For the remainder of this paper, we use $X$, $Y$, and $Z$ to denote the number of individuals in the $S$, $I$, and $R$ compartments, respectively, to avoid any confusion with $\rr$.
\begin{align}
\systeme{\frac{dX}{dt} = -\frac{\beta XY}{N}, \frac{dY}{dt} = \frac{\beta XY}{N} - \gamma Y, \frac{dZ}{dt} = \gamma Y}. \label{eq:sir}
\end{align}
In words, susceptible individuals become infected at a rate that is proportional to the percentage of infected individuals multiplied by $\beta$, the infection rate, and the number of susceptible individuals.  Infectious individuals recover at a rate of $\gamma$ multiplied by the number of infected individuals.   Whenever we refer to the ``original SIR model'' in this paper, we mean the set of equations given in Eq. \eqref{eq:sir} along with known initial state counts, $(X(0), Y(0), Z(0))$.

\noindent 
An outbreak occurs if the rate of change of infectious individuals is positive, $\frac{dY}{dt} > 0$, or equivalently, $\frac{\beta}{\gamma} > \frac{N}{X}$, since the sum of the derivatives is zero as the population $N$ is constant. That is, an outbreak occurs if  the rate of new infections is greater than the rate of recovery.  So as long as the number of initially susceptible individuals is large compared to the total population, i.e. $\frac{X}{N} \approx 1$, then an outbreak will occur if $\rr >1$, where
\begin{align*}
  \rr \overset{def}{=} \frac{\beta}{\gamma}.
  \end{align*}
  \textbf{The stochastic model}. To incorporate randomness into the model, we add noise to the compartments, namely,
  \begin{align}\label{eq:sir-noise}
    \hat{X}(t) &= X(t) + \epsilon_{X,t}\\
    \hat{Y}(t) &=  N - \hat{X}(t) - \hat{Z}(t)\nonumber\\
    \hat{Z}(t) &=  Z(t) + \epsilon_{Z,t}. \nonumber
  \end{align}
The ``hats''  in Equation \ref{eq:sir-noise} are used to distinguish random variables from the ``true,'' deterministic model without hats (e.g. $\hat{X}$ vs. $X$).  We are assuming the observations are generated from the  deterministic ODEs (possibly converted to discrete time) presented in Equation \ref{eq:sir} along with time and compartment dependent noise $\epsilon_{X,t}$ and $\epsilon_{Z,t}$.  Since $N$, the total population, is constant, then $\hat{Y}$ is adjusted accordingly.  An advantage to adding noise to the $X$ and $Z$ compartments as opposed to the $X$ and $Y$ or the $Z$ and $Y$ compartments is that we can more readily enforce the $X$ and $Z$ compartments to be monotonic in $t$.  This is useful as we expect the number of susceptible to be non-decreasing and the number of recovered to be non-increasing in an SIR disease-transmission process.

In summary, when we discuss estimators for $\rr$ for the SIR model, we mean to say we are forming an estimator of $\rr$ from the given set of data in Eq. \ref{eq:sir-noise},
\begin{align}\label{eq:data}
  \textnormal{Data} &= \left \{\left (\hat{X}(t)=x(t), \hat{Y}(t)=y(t), \hat{Z}(t)=z(t) \right ) : t=t_0, t_1, \dots, t_T\right \}, \\
  \widehat{\rr} &= m(\textnormal{Data}), \nonumber
\end{align}
where $m$ is a function of the data. We note that our SIR formulation does not include other population processes, such as vital dynamics (e.g. birth and death rates).  This will be true throughout our analysis, and we will discuss the implications of this in a later section.

\textbf{Review of \wxxsir methods}. In the following subsections, we review \wxxsir methods to estimate $\rr$, that are all of the form shown in Eq. \eqref{eq:sir-noise}, that is, an estimate of $\rr$ that is a function of XYZ data.  We examine both unspecified distribution and specified distribution-based estimates.  By unspecified-distribution, we refer to estimates that do not explicitly identify the distributions of random processes.  For example, we describe a method that estimates $\rr$ through minimizing the $L_2$ norm of the data and expected model estimates.  In contrast, distribution-based estimates refer to the methods that do explicitly identify a generative model to incorporate noise.  For example, we describe an estimate based on the chain Binomials described by \cite{abbey1952}, which describes the transmission process via Binomial draws.  

In the below sections, we describe the \wxxsir methods, but before we describe the methods, we would like to comment on some of the methods we omitted from this analysis.  Originally, we included 19 unique methods to estimate $\rr$.  Many of these 19 methods were similar enough to one another to be combined into one group.  For example, some methods used a Binomial variable to model the transmission of new infections whereas others used a Poisson variable but were otherwise the same.  A number of these similar methods were hierarchical models where the initial transmission and recovery were modeled with parameters and hyperparameters were then placed upon those parameters.  While we recognize the importance and utility of such models, we decided to focus on non-hierarchical models and consider these other models to be variants on these simplified, core models.

Other omissions were made because upon close inspection, the method presented did not fit the SIR paradigm.  For example, branching processes such as the Galton-Watson process (as seen in \cite{farrington2003}) are of historical importance in epidemiology.  However, such processes presume a limitless population size, and the recovery process is not explicitly modeled.  Despite this, many parts of the branching process can be seen in our Markov Chain model.

Another notable omission is that of the individual-based (IBM) or agent-based model (ABM).  In these models, individuals or agents interact with each and their environment over time to produce difficult-to-calculate effects \citep{blower2004,epstein2007agent}.  However, IBMs and ABMs rarely satisfy our assumption of homogeneity among individuals in the same compartment, which we require in the \wxxsir methods.  We hope to study $\rr$ estimates in such models in future work, because IBMs and ABMs present a fairly unique way to estimate $\rr$, as we can directly count who infected whom at each time point.

\subsection{Exponential Growth (EG)}\label{sec:methods-eg}

\cite{wallinga2007generation} report that the effective reproduction number $\rr$ and hence the initial reproduction number $\rr$ may derived using the fact that infection ``counts increase exponentially in the initial phase of an epidemic.''  We then have to estimate $r$, the \textit{per capita} change in the number of new cases per unit of time and $\omega$ the serial interval, the distribution of time between a primary and secondary infection. Then, we have
\begin{align}\label{eq:lotka}
\rr = \exp{(r \omega)}
\end{align}
Equation \eqref{eq:lotka} is derived from a demographic view using the Lotka-Euler survival equations which come from the fields of demography, ecology, and evolutionary biology.  Sometimes, in epidemiology, the first order Taylor series expansion of Eq. \eqref{eq:lotka} is instead used. This method is often taught as a way to introduce the concept of $R_0$ and a variant can be seen in \cite{nishiura2010b}.  Commonly, we assume that the serial interval $\omega$ is the same as infection duration,     $\omega \approx \gamma^{-1}$ and during the `initial growth phase' that the number of susceptibles is equal to
\begin{align*}
    X(t) &= X(0) e^{-r t}.
    \end{align*}
To estimate $\rr$, we then must estimate r,
\begin{align}\label{eq:exp-growth}
 \hat{r} &= \arg \min_{r} \sum_{t=t_{\min}}^{t_{\max}} \left ( X(t) - x(t) \right )^2 \nonumber\\
    \widehat{\rr} &= 1 + \frac{\hat{r}}{\gamma}.
\end{align}

The advantages to this method are that it relies only on estimates of the number of susceptibles.  However, it assumes exponential growth during the initial growth phase and so relies on knowing when that initial growth phase occurs.  \cite{nishiura2010b} provide guidelines of when such a method should be used because of the initial growth assumption.  This method also relies on having some estimate of $\gamma$ available. Tuning parameters for this method include $t_{\max}$, the maximum time point of exponential growth in the number of infections is observed (or the corresponding exponential decay in the number of susceptibles), possibly $t_{\min}$, the minimum time point, and $\gamma$ (if it is not treated as a random variable).  We choose that $t_{\min} =0$ and  $t_{\max}$ is the time corresponding to the maximum number of infectious of the observed data.

There are a number of adjustments one can make to this method.  For example, \cite{wallinga2007generation} detail a way to estimate $\rr$ through treating $\omega$ as a random variable.  Some variations of this method assume that $r$ has its own distribution \citep{obadia2012r0}.

\subsection{Ratio Estimator (RE)}\label{sec:methods-re}

Our second approach to estimate $\rr$ in the SIR model is to minimize the joint mean square error for the data collected at each time point and use the plug-in estimator found in Equation \ref{eq:sirls}.  In particular, we find:

\begin{align*}
(\hat{\beta}, \hat{\gamma} )&=\textnormal{arg} \textnormal{min}_{\beta, \gamma} \sum_{t} \left [ \left (x(t) - X(t;\beta, \gamma)\right )^2 + \left ( y(t) - Y(t;\beta, \gamma) \right )^2  + \left ( z(t) - Z(t;\beta, \gamma) \right )^2 \right ]
\end{align*}
Then the ratio estimator (RE) estimate for $\rr$ is given by Equation \ref{eq:sirls},
\begin{align}\label{eq:sirls}
  \widehat{\rr}= \frac{\hat{\beta}}{\hat{\gamma}}.
\end{align} 
The estimates resulting from this method are often found via a grid search of $\beta$ and $\gamma$ values to minimize the sum square errors.  The estimates can also be found using more sophisticated optimization algorithms. Since we typically cannot explicitly write down the partial derivatives of $S$, $I$, and $R$ with respect to $\beta$ and $\gamma$, we can only use minimization algorithms which do not rely on an explicit gradient, such as Nelder-Mead \citeyear{nelder-mead1965} simplex minimization.   RE and similar variants have been used to estimate $\rr$ in \cite{majumder2016}. 

A reasonable question one may ask is why we use the $L_2$-norm as opposed to $L_1$ or some other similarity score.  One answer is that if we were to assume independent and identically distributed Gaussian noise, then the $L_2$-norm would be equivalent to the maximum likelihood estimation.  Another answer is that $L_2$ is a continuously differentiable function and hence is easier to compute, which allows for sensitivity analysis to  be conducted more easily.  Finally, the RE estimate can be found without writing down explicit assumptions about the noise within the model.  However, we cannot guarantee properties of consistency or convergence to a known distribution without more explicit assumptions.  


For the most part, there are not many adjustments that can be made to this method because we do not explicitly specify a stochastic model for how the data are generated.  As such, we have no nuisance parameters or tuning parameters that we need to adjust when fitting our model.

\subsection{Reparameterized Ratio Estimator (rRE)}\label{sec:methods-rre}

In the previous method (Section \ref{sec:methods-re}), we estimated $\beta$ and $\gamma$ and then estimated $\rr$.  However, it is possible to directly estimate $\rr$ if we reparameterize the ODEs in Equation \eqref{eq:sir} directly with \(\rr\) and \(\gamma\), using the relation $\rr = \frac{\beta}{\gamma}$.
We find
\begin{align*}
(\widehat{\rr}, \hat{\gamma} ) &= \text{argmin}_{\rr, \gamma} \sum_{t} \left [ \left (x(t) - X(t; \rr, \gamma)\right )^2 + \left ( y(t) - Y(t; \rr, \gamma) \right )^2  +\left ( z(t) - Z(t; \rr, \gamma) \right )^2 \right ]
\end{align*}
We use the $\widehat{\rr}$ directly from the above estimation problem, which again can be found with a grid search or another optimization process.  It may be surprising that this method can lead to different results than simply using the ratio estimator.  In our simulation study, we show that we have differences in both the mean and variance of the two estimates of $\rr$.  The difference in mean can be attributed to the practical issue of finding the arguments that minimize the objective function (e.g. $R_0$ is typically not near the boundary of zero whereas $\beta$ and $\gamma$ are more likely to be closer to that boundary).  The difference in variance is due to both the difference in mean (as the mean estimate is used in the variance estimate) and also because RE estimates the variance by first estimating the Hessian of the `true' function numerically and then applying the delta method to estimate the variance of $\rr$.  On the other hand, in rRE only the Hessian needs to be estimated. 

 Like RE, while we have no explicit assumptions on the noise. Likewise, the same difficulties in sensitivity analysis arise as for the RE.  We are unaware of any attempts in the literature to directly estimate $\rr$ with rRE as opposed to estimating $\beta$ and $\gamma$ in RE and then taking the ratio to estimate $\rr$.

\subsection{Log Linear (LL)}\label{sec:methods-ll}

 \cite{harko2014exact} reduce the SIR model with two ODEs and one constraint to one ODE and one constraint.  From this, we can derive the following, 
\begin{align}
 \textnormal{log} \left ( \frac{X(t)}{X(0)} \right ) =  -\rr \frac{Z(t)}{N}. \label{eq:harko_lin}
\end{align}

We can regress the left hand side in Eq. \ref{eq:harko_lin} on $Z(t)$, the number of recovered individuals, using least squares to estimate the coefficients with $\rr$ as the coefficient of $Z(t)/N$ to obtain an estimate of $\rr$,
\begin{align*}
  \widehat{\rr} = -\frac{\sum_{t=0}^T \textnormal{log} \left (\frac{  X(t)}{X(0)} \right )}{\sum_{t=0}^T\frac{Z(t)}{N}}.
\end{align*}
This method says that for a one percent increase in the number of recovered individuals over time period $\Delta$, we expect the ratio of the number of previous susceptibles to number of new susceptibles to increase by $e^{.01\rr}$,
\begin{align*}
  \frac{X(t)}{X(t+\Delta)}  &= e^{.01\rr}.
\end{align*}

An advantage of this method is an alternative interpreation of $\rr$: in terms of the ratio of the old number of susceptibles to new number of susceptibles for a specific amount of time.  Another advantage is that  this method is easily implemented using any linear modeling software.  To our knowledge, the log-linear model has not yet been applied to a data setting.

The method also has disadvantages in that it is assumes the expected path of the individuals exactly follows the Kermack and McKendrick differential equations, which makes this model choice very sensitive to the disease transmission process.  As a result, we currently do not know of any variants for this method.

\subsection{Markov Chain (MC)}\label{sec:methods-mc}

A natural approach to epidemic modelling is that of Markov chains (MC), since it is assumed an individual's next state is only dependent on its current state and the current states of other individuals.  Much work has been done over the years in this specific field including asymptotic behavior, continuous time MC, confidence intervals, and more \citep{jacquez1991,gani1995,daley2001epidemic}.  We present one simple instance of the model, the discrete time case, which traces its origin back to the Reed-Frost model \citep{abbey1952}.

In this model, the number of susceptibles at the next step, $\hat{X}(t+1)$, has a Binomial distribution based on the contacts with the current number of infectious, $\hat{Y}(t)$ and the current number of susceptibles.  This is based on estimating $\alpha$, the probability of coming into contact and subsequently becoming infected by a single infectious individual.

Unlike some of the other methods, the Reed-Frost Chain Binomial has a very specific form of $Y(t)$, the number of infectious at time $t$.  We assume that that the incidence at time $t$ is equal to the number of infectious at at time $t$,
\begin{align*}
    \hat{X}(t) &= \hat{X}(t-1) - \hat{Y}(t)\\
    \hat{Y}(t)|\hat{X}(t-1), \hat{Y}(t-1) &\sim \textnormal{Binomial}\left (\hat{X}(t-1), 1 - (1-\alpha)^{\hat{Y}(t-1)}\right )\\
    \hat{Z}(t) &= \hat{Z}(t-1) + \hat{Y}(t)
\end{align*}
 The full likelihood for $\alpha$ in this model, namely,
\begin{align*}
\mathcal{L}\left ( \alpha ; \textnormal{Data}\right ) \propto \prod_{t=0}^{T-1}\left ( (1- \alpha)^{\hat{Y}(t)} \right )^{\hat{X}(t+1)} \left (1-(1- \alpha)^{\hat{Y}(t)} \right )^{\hat{X}(t) - \hat{X}(t+1)}.
\end{align*}
Maximizing the likelihood yields an estimate for $\alpha$,
\begin{align*}
\hat{\alpha} = \textnormal{arg} \textnormal{max}_{\alpha} \mathcal{L}\left(\alpha; \textnormal{data} \right ).
  \end{align*}
An approximate estimate of the reproduction number is then,
\begin{align}\label{eq:r0-mc}
\widehat{\rr} &= \textnormal{log} \left ( \frac{1}{1-\hat{\alpha}}\right ).
\end{align}

This method typically allows for more than just the reproduction number to be estimated.  Through recursion, one can estimate the probability of having a given number of susceptibles and infected at each time step, and hence the entire probability distribution may be known. The advantages of this model are its simplicity of interpretation and ability to generate an estimate for the probability distribution for $(X(t), Y(t))$. 


Essentially, in this model we assume that all infectious cases have recovered by the next time step or equivalently, that $\gamma = 1$.  As a result, it is likely that this method will overestimate $\rr$ except in cases where, for example, infectiousness does not appear until symptom onset and at discovery of symptoms the patient is immediately treated and no longer infectious.

\subsection{Likelihood-Based Estimation (LBE)}\label{sec:methods-lbe}
A common way used to estimate epidemiological quantities is through likelihood based estimation.  We saw an example of this in the previous section with MC estimation.  However, as we will see in the results, this particular model does not seem to fit the data well.  As a result, many researchers specify different models for the generation of SIR data; estimate the likelihood through direct calculation or approximate the likelihood through simulation such as iterated filtering used in ``plug-and-play'' and finally derive point estimates and CIs from this likelihood estimate  $\rr$  \citep{abbey1952,forsberg2008,king2009,bhadra2012}.  Here, we examine one model that has been used to model the discrete-time SIR: a binomial chain with probabilities based on the deterministic SIR model.


The discrete time XYZ model is as follows, where we have replaced $d$ with $\Delta$,
\begin{align*}
\systeme{\frac{\Delta X}{\Delta t} = -\frac{\beta Y X}{N}, \frac{\Delta Y}{\Delta t} = \frac{\beta Y X}{N} - \gamma Y, \frac{\Delta Z}{\Delta t} = \gamma Y}. 
\end{align*} 
Denote $\hat{X}(t), \hat{Y}(t)$, and $\hat{Z}(t)$ to be the observed number in each compartment and $X(t), Y(t)$, and $Z(t)$ to be the true, underlying model at time $t$.   The stochastic model is given by
\begin{align*}
  \hat{X}(t+1) | \hat{X}(t), \hat{Z}(t) &= \hat{X}(t) - \epsilon_{X,t}\\
  \hat{Y}(t+1) | \hat{X}(t), \hat{Z}(t)& = N - \hat{X}(t+1) - \hat{Z}(t+1), \nonumber \\
  \hat{Z}(t+1) | \hat{X}(t), \hat{Z}(t)&= \hat{Z}(t) + \epsilon_{Z,t} \nonumber
\end{align*}
with $\hat{X}(0)$, $\hat{Y}(0)$, and $\hat{Z}(0)$ known. The terms $\epsilon_{X_t}$ and $\epsilon_{Z_t}$ in Equation \eqref{eq:sir-noise} incorporate random behavior into the model and we constrain $\beta, \gamma \in [0,1]$.  Here we choose them to have the following distributions,
\begin{align*}
 \epsilon_{X,t}| \hat{X}(t), \hat{Z}(t)\ &\sim \text{Binomial}\left (\hat{X}(t), \frac{\beta \hat{Y}(t)}{N} \right ) \\
  \epsilon_{Z,t}| \hat{X}(t), \hat{Z}(t)&\sim \text{Binomial}\left (\hat{Y}(t), \gamma \right ). \nonumber
\end{align*}  If we assume that the next time step is dependent only on the current time step, then the likelihood is
\begin{align*}
  \mathcal{L}\left (\beta, \gamma ; \textnormal{Data} \right ) &=
                                                  \prod_{t=0}^{T-1} \left [\left ( \begin{array}{c}
                                                           x_t \\
                                                             \epsilon_{X,t}\
                                                            \end{array}
  \right ) \left (\frac{\beta y(t)}{N} \right )^{ \epsilon_{X,t}}\left (1 - \frac{\beta y(t)}{N} \right )^{x(t) - \epsilon_{X,t}} \cdot \right .\\
  & \left . \left ( \begin{array}{c}
              y(t)  \\
              \epsilon_{Z,t} 
            \end{array}
  \right ) \gamma^{\epsilon_{Z,t} }\left ( 1 - \gamma\right )^{y(t) - \epsilon_{Z,t}} \right ].
\end{align*} The estimate for $\rr$ is then
\begin{align}\label{eq:lbe-r0}
  (\hat{\beta}, \hat{\gamma}) &= \textnormal{arg} \textnormal{max}_{\beta, \gamma} \mathcal{L}\left ( \beta, \gamma; \textnormal{Data} \right )\nonumber\\
  \rr &= \frac{\hat{\beta}}{\hat{\gamma}}
  \end{align}
  The two main differences between this model and MC are  that 1) LBE looks at the joint likelihood of the XYZ model, whereas MC assumes the previous infectious individuals all recover from one time step to the next and 2) the difference in the structure of the transition probability. A major advantage of this model easily modified and can be used to as a basis for more sophisticated models, and can be easily estimated in available software such as \texttt{pomp} \citep{king2016}.  Disadvantages include maximizing the likelihood may be more time consuming than other methods due to the need to approximate $X(t)$ and $Z(t)$ for many different values of $\beta$ and $\gamma$ and that this model is sensitive to the underlying assumptions of how the disease is stochastically transmitted.  There are no tuning or nuisance parameters that need to be estimated.

\subsection{Incidence to Prevalence (IPR)}\label{sec:methods-ipr}

The incidence to prevalence ratio (IPR), described by \cite{Nishiura2009}, is another intuitive method to estimate $\rr$.  It incorporates some of the most basic epidemiological quantities: incidence and prevalence. In terms of data from the SIR model, the incidence is approximately the number of new infectious for a given time step, $J(t) \approx -(X(t+1) - X(t))$, and the IPR is the ratio of incidence to prevalence, IPR$(t) = \frac{J(t)}{Y(t)}$.  This method assumes that we have some prior knowledge about $\gamma$, the recovery rate.  Thus we use as our estimate,
\begin{align*}
\widehat{\rr} &= \textnormal{IPR}(t^*) \cdot \frac{1}{\gamma}
\end{align*}
where $t^*$ is the time at which the maximum prevalence is reached.

Here we assume that the time step is small enough to approximate the incidence.  The advantage of this method is that incidence data is generally readily available as is prevalence data for certain diseases.  However, as one is required to have prior knowledge about $\gamma$, it may be easier to directly estimate $\rr$ with one of the many other methods described that does not require  prior knowledge.  Again, we are using only one time point to estimate $\rr$.  This model places no assumptions on the noise.  Here, $\gamma$ may be treated as a tuning parameter.

\subsection{Linear model approximation (LMA)}\label{sec:methods-lma}
In the past decade, non-parametric and semi non-parametric methods have been increasingly used to (quite successfully) predict the prevalence of a disease such as the methods by \cite{brooks2015} used in Center of Disease Control and Prevention's yearly influenza prediction contest.  This success in prediction prompted us to ask whether success in non-parametric curve fitting could transfer to estimating $\rr$.

The idea is as follows: assume $(\hat{X}(t), \hat{Y}(t), \hat{Z}(t)) = (f(t), g(t), h(t)) + \bf{\epsilon}_t$ where $f, g, h$ are functions such that $f^\prime(t)$, $g^\prime$(t), and $h^\prime(t)$ exist for all $t \in (0, T)$.  Then a rough estimate of $\rr$ is
\begin{align*}
    \widehat{\rr} &\approx -\frac{\hat{f}^\prime(0)}{\hat{h}^\prime(0)} \times \frac{N}{X(0)}
\end{align*}
where $\hat{f}$ and $\hat{h}$ are smooth estimates of the actual underlying distributions.  For example, we could have $\hat{f}$ be a polynomial function in $t$.

The linear approximation of the Kermack and McKendrick SIR model has been studied previously \citep{chen2009,hu2014}.  We extend these linear models to polynomial models.  Alternatively, one can think of this method as a naive way to approximate the S, I, and R curves.  We use this method to estimate $\rr$. Specifically, we estimate two polynomials in \(t\) with degree $K$  using least squares to find the coefficients $\{(\hat{x}_k,
\hat{z}_k)\}_{k=1, \dots, K}$ and use the resulting polynomial coefficients to estimate the derivatives, $\hat{X}^\prime(t)$ and $\hat{Y}^{\prime}(t)$.
An estimator for \(\rr\) is derived from the ODEs in Equation \eqref{eq:sir},
\begin{align}\label{eq:r0-lma}
  \widehat{\rr} &= -\frac{\hat{X}^\prime(0)}{ \hat{Z}^\prime(0)} \cdot \frac{N}{\hat{X}(0)}. 
  \end{align}
  Here, $K$ is arbitrary and should be selected using some criterion such as AIC.  Besides optionally deciding on the degree of polynomials to fit, this model is simple to implement.  The time $t=0$ is used to best capture the initial outbreak. An advantage of using this estimation method is that it is simple to implement using any linear modelling software.

A prominent disadvantage of this method (and in general using semi or non-parametric curve estimates to estimate $\rr$) is that we implicitly assume that the underlying disease curves are mechanistic in nature, specifically of the mechanism of disease transmission indicated by the SIR model. As such, it may make no sense in the first place to estimate the the curves in a non-parametric manner, when ultimately we are using those estimates to estimate a disease parameter. Moreover, if the disease transmission process is not truly described by the SIR model, then our derivation of $\rr$ may be meaningless and so we may be estimating the wrong quantity of interest. 

Despite this prominent disadvantage, we include this model in our testing because non-parametric methods have been successfully used recently for the prediction of disease models, and it is a natural question of whether we can `plug and chug' the curve estimates to obtain a reliable estimate of $\rr$. If the estimates are reliable (or reliable enough), then we can use curves for both prediction and inference.

\textbf{Methods Summary}. Sections \ref{sec:methods-eg}-\ref{sec:methods-lma} describe how to form an estimate of $\rr$ for \wxxsir different methods.  The \wxxsir methods are summarized in Table \ref{tab:methods}.  All of our methods explicitly describe point estimates of $\rr$.  In addition, we report 95\% CI to go along with every point estimate.  These 95\% CI are determined by appropriate use of the Delta Method as described in \cite{wasserman2004}.
\begin{table}
\centering
\caption{Table of methods used.  The heading ``Specified distribution?'' refers to whether a statistical distribution is explicitly specified in the model.  The methods are described in Sections \ref{sec:methods-eg}-\ref{sec:methods-lma}.}
\label{tab:methods}
\resizebox{\textwidth}{!}{%
\begin{tabular}{@{}llcl@{}}
\toprule
\textbf{Abbr.} & \textbf{Method}             & \textbf{Specified distribution?} & \textbf{Reason for inclusion}                   \\ \midrule
EG             & Exponential Growth          & No               & Naive approximation \\
RE             & Ratio Estimator             & No                 & History, simplicity, optimization \\
rRE            & Reparamatrized RE           & No                  & Slight adjustment to known model  \\
LL             & Log Linear                  & No                   & Slight adjustment to known model  \\
MC             & Markov Chain                & Yes              & History, plausibility             \\
LBE            & Likelihood Based Estimation & Yes                  & Common method of estimation       \\
IPR            & Incidence to Prevalence     & No                   & Typical data                      \\
LMA            & Linear Model Assumption            & No              & Non-parametric approach          \\ \bottomrule
\end{tabular}%
}
\end{table}

\section{Data}\label{sec:data}
An important and sometimes glossed-over issue in estimating parameters in SIR models (and compartment models in general) is how the data is ``pre-processed'' into a format that the mentioned methods can use.  The methods for estimating $\rr$ in Section \ref{sec:methods} rely on observations of $(x(t), y(t), z(t))$, the number of individuals in the susceptible, infectious, and recovered compartments at different time points.  However, such observations only occur in fairly contrived situations such as small, closed populations where infection time and recovery time are almost immediately known.  What is far more common is, instead, to have a fairly good estimate of the population size $N$ and the new case counts (incidence) $j(t)$ on a daily, weekly, or monthly basis.  From there, the data may be pre-processed into a usable form that the above methods may use.  For example, if $N$ is large, the number of susceptibles can be reasonably estimated from the case counts as $x(t) = N - \sum_{s=0}^t j(s)$, the population size minus the cumulative sum of the incidence over the time period.  However, it is typically more difficult to disentangle $y(t)$ and $z(t)$.

One approach  to separate the number of infectious from the number of recovered  is to employ the use of a known estimate of $\gamma$, the rate of recovery or equivalently $\gamma^{-1}$, the average time to recovery for an infectious individual, as seen in \cite{towers2009}.  We assume that the number of infectious is  the previous number of infectious, plus the new cases and minus the new recovered cases.  In fact, the above is a discrete approximation of the SIR model where $j(t)$ is the new cases flowing in, $y(t-1)$ is the previous number of cases and $\gamma y(t-1)$ is the number of cases flowing out of the infectious state and so are subtracted.  The transformation from incidence to the XYZ values are given by the following equations,
\begin{align}\label{eq:inc-to-xyz}
x(t) & =  N - \sum_{s=0}^t j(s) \nonumber \\
    y(t) &= j(t) + y(t-1)(1 - \gamma) \nonumber \\
    z(t) &= N - x(t) - y(t) ,
\end{align}

The consequence of pre-processing the data is that we are imposing SIR (XYZ) structure on our observed data.  This will almost certainly have an effect of estimating $\gamma$ and thus $R_0$, but this pre-processing step remains a common approach (as seen recently in \cite{wang2020}).  Other approaches to pre-processing include treating the XYZ compartment values as latent variables or through data augmentation \citep{king2009,fintzi2016}.  We find these other approaches to be useful and valuable but do not analyze them here.

In our simulation study, we generate and fit our methods on data of the form ($x(t)$, $y(t)$, $z(t)$), i.e. the ideal situation where we know the counts of all three compartments.  In contrast, the 2009 pandemic influenza data uses an approximation of the incidence (weighted influenza like illness).  In that case, we use the pre-processing steps described above in Eq. \ref{eq:inc-to-xyz} to format the data into a usable SIR format.  Because the data is approximated, we perform sensitivity analysis on $\gamma$.

\subsection{Simulation data}

In this section, we give the details of how we generate our simulation data.  Since we want to compare the \wxxsir methods to one another, we strive to compare data with both process and measurement error, where the former refers to the error from the inherent variability in transmission model, and the latter to observation error. In all of these simulations, we first generate data from the models under known functions ($f$, $g$). We then add noise drawn from known distributions $(\epsilon_{X,t}, \epsilon_{Y,t})$ to each time point,
\begin{align}\label{eq:sim-models}
  \hat{X}(t) &= f(t) + \epsilon_{X,t} \\
  \hat{Y}(t) &= N - \hat{X}(t) - \hat{Z}(t) \nonumber\\
  \hat{Z}(t) &= N  + \epsilon_{Z,t}.\nonumber 
\end{align}
To maintain a constant population, we adjust the recovered compartment $\hat{Y}(t)$ after $\hat{X}(t)$ and $\hat{Z}(t)$ are determined.  We simulate data that has a combination of process error and measurement error.  We assume that the initial values $(\hat{X}(0), \hat{Y}(0), \hat{Z}(0))$ are known.

\noindent \textbf{Simulation data generation}
\begin{align}\label{eq:par}
  f(t) &= \hat{X}(t-1)\nonumber\\
  g(t) &= \hat{Z}(t-1)\nonumber\\
    \epsilon_{X,t} | \hat{X}(t-1), \hat{Z}(t-1)&= \alpha_{X,t} + \tau_{X,t}\nonumber\\
        \epsilon_{Z,t} | \hat{X}(t-1), \hat{Z}(t-1)&= \alpha_{Z,t} + \tau_{Z,t}\nonumber\\
     \alpha_{X,t}| \hat{X}(t-1), \hat{Z}_{PAR}(t-1) &\sim \textnormal{Binomial} \left ( \hat{X}(t-1), \frac{\hat{Y}(t-1)}{N} \right ) \nonumber\\
     \alpha_{Z,t}| \hat{X}(t-1), \hat{Z}(t-1) &\sim \textnormal{Binomial} \left ( \hat{Y}(t-1), \gamma \right ) \nonumber \\
     \tau_{X,t} & \sim N\left(0,  \sigma^2_{X}\right )\nonumber\\
     \tau_{Z,t} &\sim N\left(0,  \sigma^2_{Z}\right)
\end{align}
The process error in our simulations is shown through the Binomial draws, which noisily approximate the Kermack and McKendrick SIR equations.  Measurement error is added through autoregressive error which can compound over time.  In this simulation, we set $R_0$ = 2, which is an upper bound estimate of $\rr$ for pandemic influenza.  We let the total population size be $N=100,000$, approximately the size of an average US county.  The rest of the parameter specifications for the baseline simulation data are shown in Table \ref{tab:baseline-params}.
 \ref{tab:baseline-params}.

  \begin{table}[H]
      \caption{Parameter values for baseline simulated data.  The simulation data is generated in accordance to the model in Eq. \eqref{eq:par}.}
    \label{tab:baseline-params}
\centering
      \begin{tabular}{@{}lllllllllll@{}}
        \toprule
        \textbf{Parameter} & $\beta$ & $\gamma$ & $\rr$ & $T$ & $X(0)$ & $Y(0)$ & $Z(0)$ & $\sigma_X$ & $\sigma_Z$  \\ \midrule
        \textbf{Value}     & 0.02    & 0.01     & 2.00  & 365 & 99950  & 50     & 0      & 100        & 5           \\ \bottomrule
      \end{tabular}%

  \end{table}

We examine three variations to the baseline data:  changes in time points used $(T)$, susceptible to initial infectious ratio ($X(0):Y(0)$), and total population size $N = X(0) + Y(0) + Z(0)$.  Additionally, we examine two data sets where non-Kermack and McKendrick SIR models are the true generative models.  The first data set is two independent SIR models (e.g. children and adult infections) that are initialized with different $\rr$ values (1.5 and 2.5) and the final data is accumulated together (e.g. $X_1(t) + X_2(t) = X(t)$).  The second data set is an SEIR (XEYZ) model with $\rr = 2$ such that the $X$ and $E$ compartments are combined together (i.e. exposed individuals are observed to be susceptible).  More details of how these non Kermack and McKendrick SIR data sets are simulated are described in Appendix \ref{app:sim-data}.

\subsection{2009 Pandemic influenza data}
   The data for the 2009 pandemic influenza are publicly available from the Center of Disease Control and Prevention's (CDC) FluView \citep{cdc-fluview}.  The data source is ILINet and is the national level data.  We use a similar date range as in \cite{towers2009} who use data from epiweeks 21-33 (May 23, 2009-August 22, 2009).  This time range turned out to be the first wave of H1N1 in the USA.  The disease later had a more severe outbreak, but we do not analyze that period here.

      From the United States Census Bureau, we use the population estimate of the USA as of April 1, 2010 of $N=308,740,000$ individuals \citep{census-2010}.  The main features in this data set are weighted influenza like illness (wILI) for a given week (Epiweek).  The wILI values are estimated based on the number of patients diagnosed with influenza like illness and weighted based on the reporting locations.  To obtain weekly incidence from wILI, we use $J(t) = \textnormal{wILI}(t) / 100 \cdot N$. 
      
\begin{figure}
  \centering
  \includegraphics[width=\textwidth]{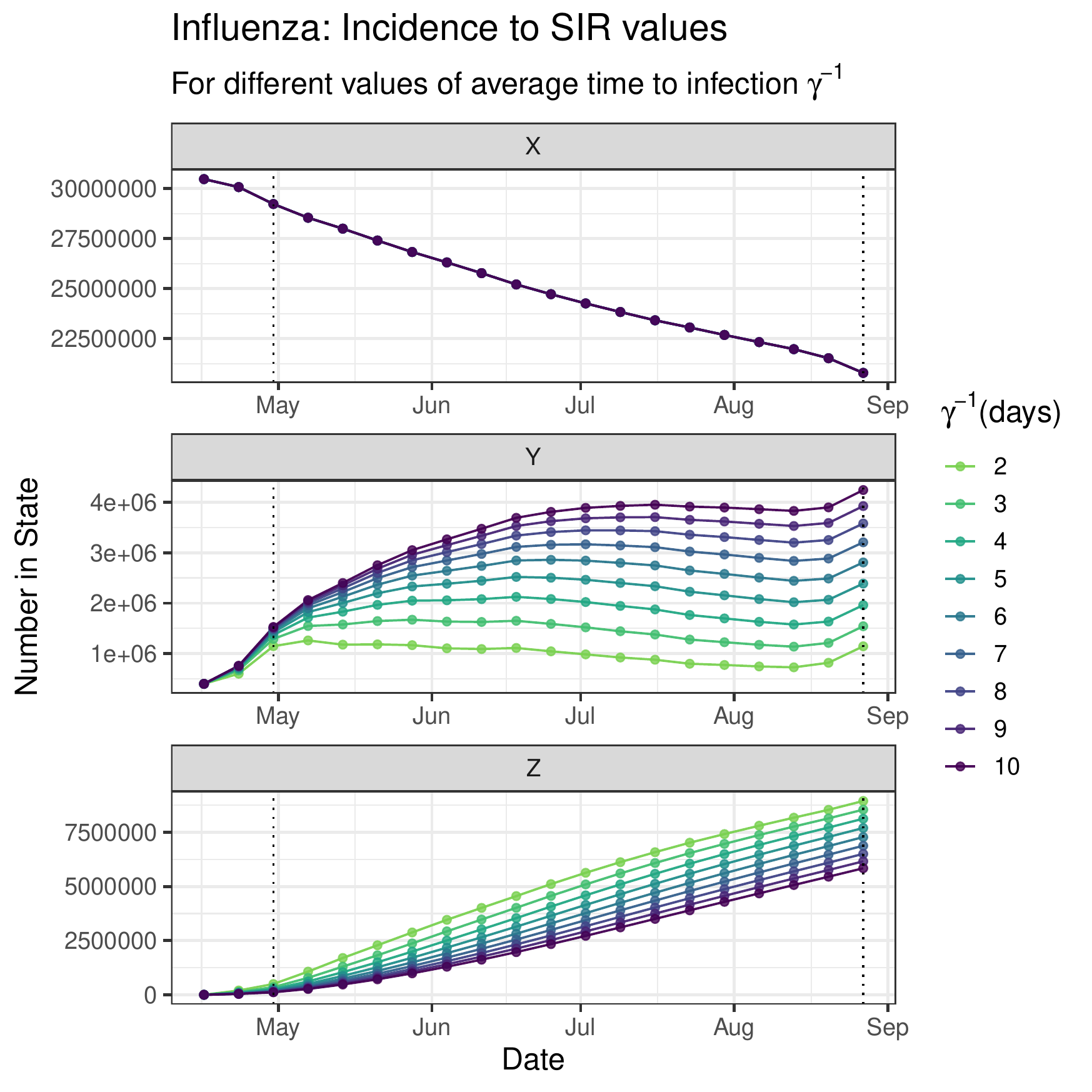}
  \caption{Observed data from H1N1 influenza 2009 in the USA.  Weekly wILI is reported, which we transform into S, I, and R counts shown for weeks April 16-Aug. 27.  We see large variation in both the number and the shape of the Y and Z compartments for values between 2 and 10 days of the average time to recovery $(\gamma^{-1})$.}\label{fig:h1n1-data}
  \end{figure}

      To impute the number of individuals in each state at different time points, we use the relations described in Eq. \eqref{eq:inc-to-xyz}, which requires an estimate of $\gamma^{-1}$, the average expected time from infection to recovery. We use $\gamma^{-1} = 3$ days which was estimated by \cite{vespignani2007} and was also the estimate used by \cite{towers2009}.  Because $\gamma^{-1}$ is a random variable, we perform sensitivity analysis and report $\rr$ estimates for multiple values of $\gamma^{-1}$. The data is summarized in Table \ref{tab:h1n1-data} and the pre-processed data is shown in Figure \ref{fig:h1n1-data}.  Looking at the pre-processed data with differing values of $\gamma^{-1}$, we see that the course of the outbreak can drastically differ depending on the average infectious period.

\begin{table}[H]
\caption{Summary of the data used to estimate $\rr$ during the 2009 H1N1 influenza pandemic in the USA.  We follow the data collection and SIR imputation procedure described in \cite{towers2009} (although we begin our analysis on April 16 instead of May 23) and shown in Eq. \eqref{eq:inc-to-xyz}.}
\label{tab:h1n1-data}
\centering
\begin{tabular}{@{}ll@{}}
\toprule
Feature       & Notes                                                                   \\ \midrule
Disease       & H1N1 Pandemic Influenza A                                                 \\ 
Source        & CDC FluView (\url{https://www.cdc.gov/flu/weekly/})      \\
Year          & 2009                                                                    \\
Date(s)          & April 16-August 22 (Epiweeks 15-33)                                       \\
Location      & USA                                                                     \\
Population    & $N$=308,740,000 people                                                  \\
  Avg. Duration $\left ( \gamma^{-1}\right )$ & 3 days (and sensitivity analysis)\\
  Raw data & Weekly wILI reports\\
  Imputed data & X$(t)$, Y$(t)$, Z$(t)$ \\\bottomrule
\end{tabular}

\end{table}

\section{Results}\label{sec:results}

\subsection{Simulation Results}
We present the results of a single, arbitrary simulation for each specified set of initial parameters. The reason for this is 1) because epidemic data typically occurs within a single data set and 2) so readers see how the point and CI estimates look together.  As a note, we ran thousands of simulations that are not shown here but are available for reproduction in our code repository at \url{github.com/skgallagher/r0}.  The results described and shown here are representative of our simulations as a whole.

\begin{figure}
    \centering
    \includegraphics[width=.7\textwidth]{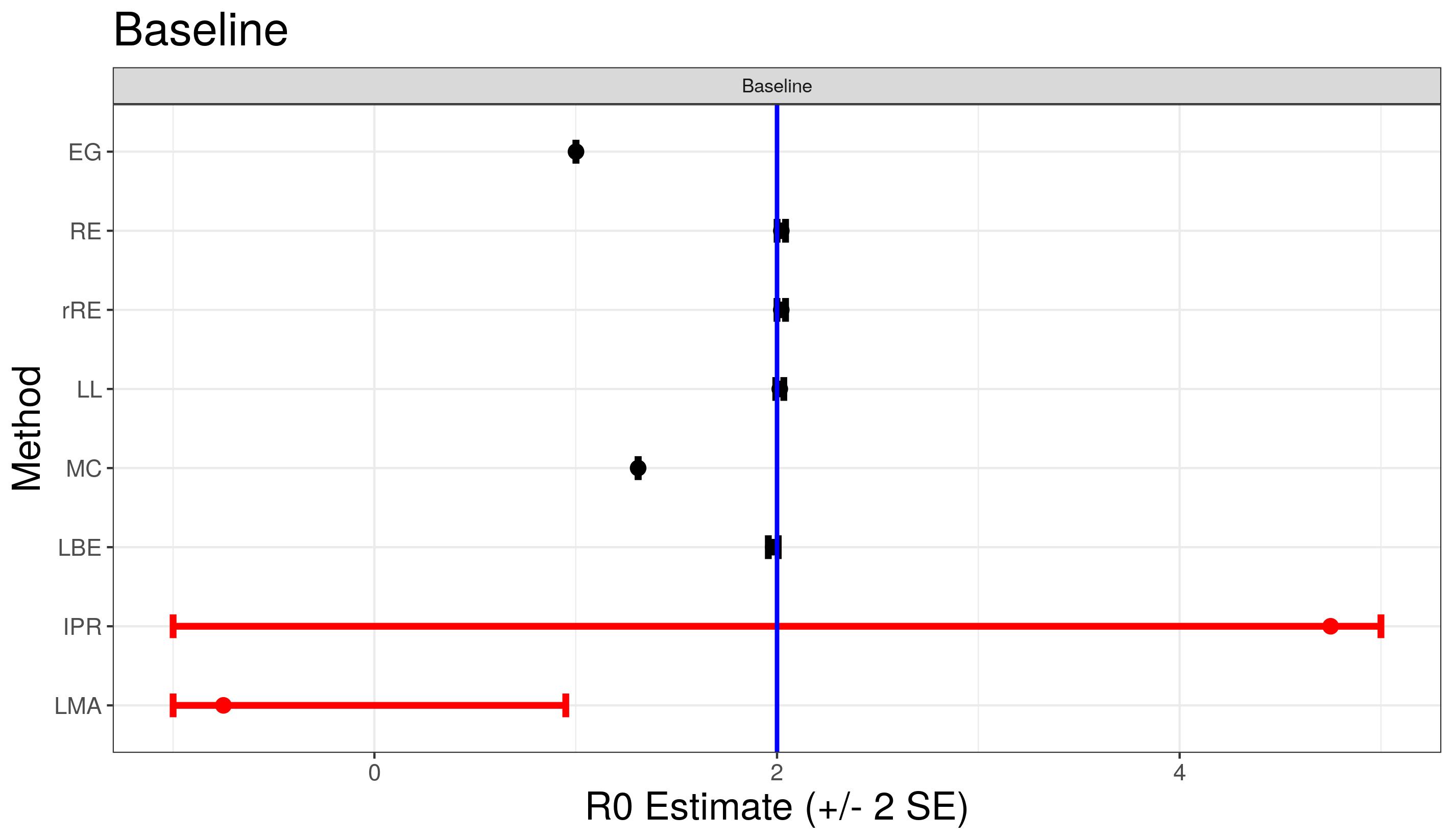}
    \caption{Point and 95\% CI estimates of a result of a single simulated data set for the given parameter conditions.  The vertical blue line denotes the true value of $\rr$.  The estimates colored in red have CIs that are truncated at least on one side for visibility.  The corresponding table of estimates are shown in Table \ref{tab:baseline-res}.}
    \label{fig:baseline-res}
\end{figure}

We find the estimates from RE, rRE, LL, and LBE to be accurate and informative (in terms of CI size) and the estimates of of EG, MC, IPR, and LMA to be both inaccurate and uninformative.  Point and 95\% CI estimates for the baseline simulated data are shown in Figure \ref{fig:baseline-res} and the corresponding table of numbers are shown in Table \ref{tab:baseline-res}.  These results indicate that both 1) a variety of methods can be used to reach similar conclusions such as RE, rRE, LL, LBE all indicate that $\widehat{\rr} \approx 2$ and 2) that naive application of methods can lead to poor interpretations of $\rr$ (EG, MC, IPR, LMA).  For example, we see that exponential growth is not a good approximation of the Kermack and McKendrick SIR model and should not be used to obtain `back-of-the-envelope' estimates of $\rr$.

In our simulations, we find that the unspecified-distribution methods of RE, rRE, and LL produce accurate and informative estimates of $\rr$, which may be especially useful when the specifics of disease transmission are unknown.  On the other hand, our likelihood based method of LBE, a fairly simple model of disease transmission, produces comparable results to that of RE, rRE, and LL., Since the basic LBE model is an accurate and informative estimate, it can used to verify more complex likelihood based models where more is known about the disease transmission process.

Note that the successful estimates closely model the true transmission function, especially in comparison to EG, MC, and LMA which make considerable approximations to the Kermack and McKendrick SIR model.  IPR is an outlier in that although it closely models the transmission function, it produces estimates that are neither reliable or informative.

\begin{table}
\caption{\label{tab:baseline-res}Table of $\rr$ point estimates and standard error (SE) for the baseline simulated data. These numbers correspond to the forest plots in Figure \ref{fig:baseline-res}.}
\centering
\begin{tabular}{@{}ll@{}}
\toprule
\bf{Methods} & $\widehat{\rr}$ $(SE(\widehat{\rr}))$\\
\midrule
EG & 1.002 (1e-04)\\
RE & 2.021 (0.0105)\\
rRE & 2.021 (0.0106)\\
LL & 2.014 (0.01)\\
MC & 1.31 ($<$1e-04)\\
LBE & 1.981 (0.0123)\\
IPR & 16.482 (21.8991)\\
LMA & -0.944 (0.9473)\\
\bottomrule
\end{tabular}
\end{table}

\begin{figure}
    \centering
    \begin{minipage}{0.32\textwidth}
        \centering
        \includegraphics[width=0.98\textwidth]{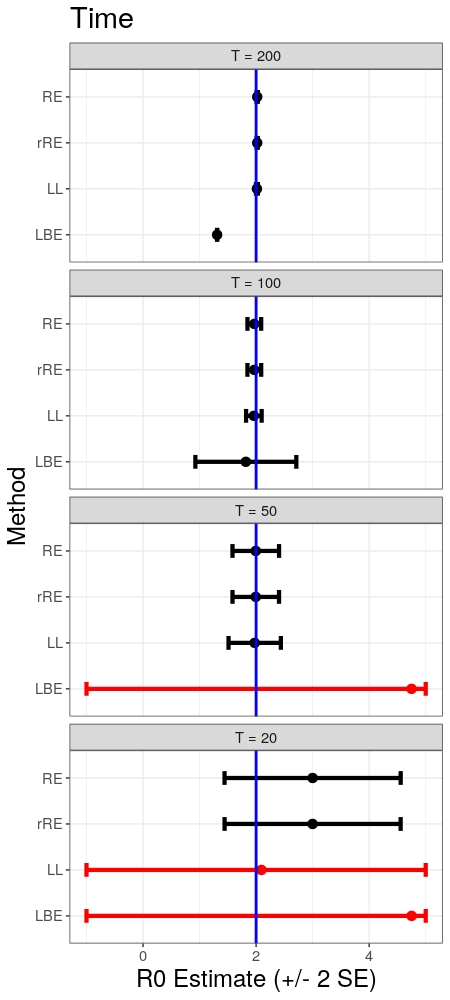} 
    \end{minipage}\hfill
    \begin{minipage}{0.32\textwidth}
        \centering
        \includegraphics[width=0.98\textwidth]{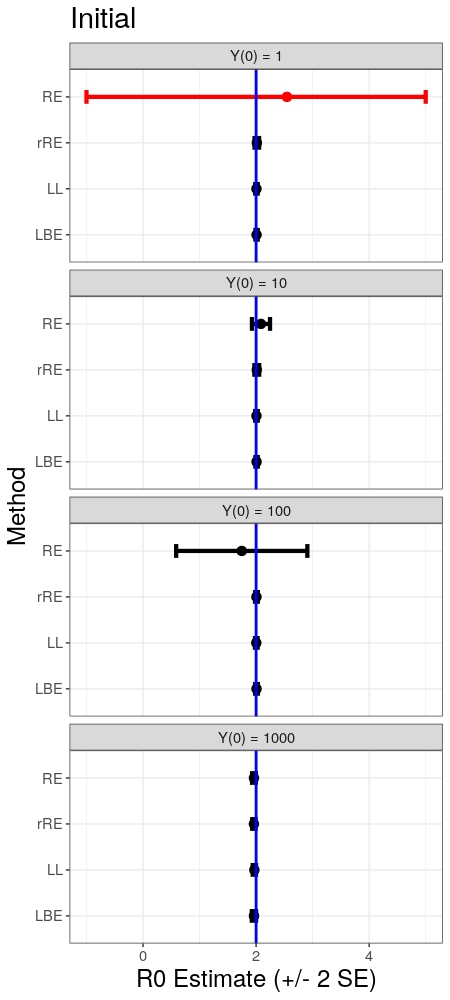} 
    \end{minipage}\hfill
    \begin{minipage}{0.32\textwidth}
        \centering
        \includegraphics[width=0.98\textwidth]{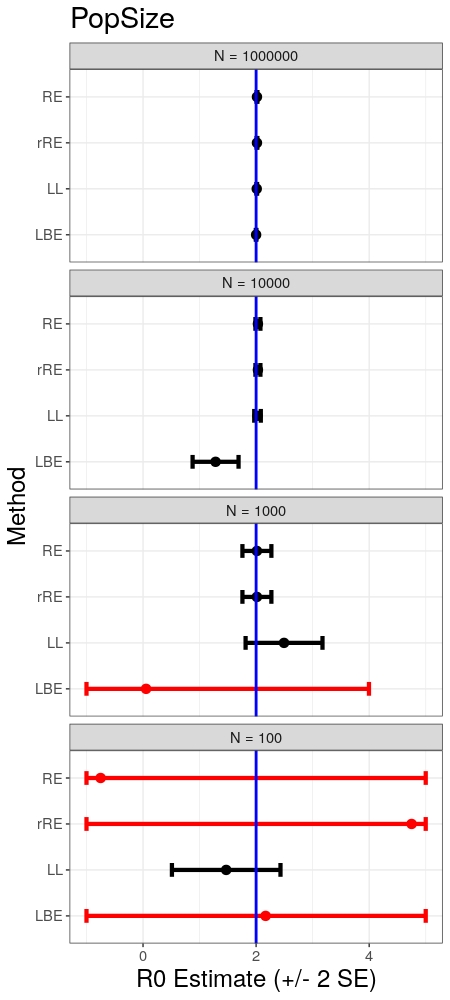} 

    \end{minipage}
    \caption{Point and 95\% CI estimates of a result of a single simulated data set for the given parameter conditions.  The vertical blue line denotes the true value of $\rr$.  The estimates colored in red have CIs that are truncated at least on one side for visibility.  The corresponding table of estimates are shown in Table \ref{tab:baseline-vars}.}\label{fig:baseline-variations}
\end{figure}

In addition to showing estimates of $\rr$ for the baseline simulated data, we also show the difference among the methods when we vary three important disease parameters, the number of time steps used $(T)$, the initial percent of infectious individuals $Y(0)/(X(0) + Y(0))$, and the population size $N = X(0) +Y(0)$.  These results show variation from the baseline simulated data in one parameter at a time and are summarized visually in Figure \ref{fig:baseline-variations} and Table \ref{tab:baseline-vars}.  We only show the results from the methods of RE, rRE, LL, and LBE so as to more closely examine differences among our more successful method estimates.

With respect to the number of time points used (left column in Fig. \ref{fig:baseline-variations}, we see that confidence intervals increase as the number of time points decreases.  This is especially apparent for LBE.  When $T=20$ (early in the outbreak), note how all methods overestimate the true value of $\rr$. With respect to the initial percent of infectious individuals (middle column in Fig. \ref{fig:baseline-variations}), for the most part the estimates of $\rr$ from RE, rRE, LL, and LBE are accurate and informative, except when there is exactly one infectious individual.  In this case, we see RE produces a very large CI.  With respect to the population size, we find that estimates decrease in accuracy and increase in CI width as $N$ decreases.  This is especially apparent when the initial population size is quite small (N=100), where none of the estimates produce an accurate estimate of $\rr$ and only LL produces a somewhat informative estimate.

In some of the estimates from LBE in Figure \ref{fig:baseline-variations}, $\widehat{\rr}$ has wider CI than we would perhaps expect or want.  Part of this is that LBE is particularly sensitive to the data generation process, namely in that it requires the number of susceptibles to be monotonically non-increasing and the number of recovered to be monotonically non-decreasing.  However, the autoregressive observation error we add in the simulation is not constrained to this (and we have observed data in the wild that also defies this constraint).  This goes to highlight the nuances and difficulties involved in these estimates.


\begin{table}
\caption{\label{tab:baseline-vars}Table of $\rr$ point estimates and standard error (SE) for simulated data that differs from the baseline simulated data by either number of time steps, population size, and initial percent of infectious. These numbers correspond to the forest plots in Figure \ref{fig:baseline-variations}.}
\centering
\resizebox{\textwidth}{!}{
\begin{tabular}{@{}llllll@{}}
\toprule
\bf{Condition} & RE & rRE & LL & LBE\\
\midrule
$T=200$ & 2.019 (0.0052) & 2.019 (0.0049) & 2.014 (0.0085) & 1.994 (0.0081) \\
$T=100$ & 1.968 (0.0607)  & 1.968 (0.0608) & 1.96 (0.0688) & 1.899 (0.0723) \\
$T=50$ & 1.994 (0.2058) & 1.994 (0.2057) & 1.975 (0.2314) & 3.068 (2.2356) \\
$T=20$ &  2.999 (0.7789) & 2.999 (0.7785) & 2.094 (1.7922) & 3.133 (2.4472)\\
\midrule 
$(X(0) = 99,999, Y(0) =1 )$& 2.545 (3.0645) & 2.086 (0.0742)  &  1.746 (0.5873) & 1.962 (0.0168)  \\
$(X(0) = 99,990, Y(0) =10 )$ &2.013 (0.0213)   & 2.012 (0.0212) &  2.006 (0.0124) &1.961 (0.0148) \\
$(X(0) = 99,900, Y(0) =100 )$&  2.006 (0.0117) &2.006 (0.0118)  & 2.003 (0.0117)  & 1.969 (0.0146)  \\
$(X(0) = 99,000, Y(0) =1,000 )$& 2.008 (0.0101)& 2.008 (0.0102)&  2.007 (0.0126)& 1.962 (0.0186) \\
\midrule 
$(X(0) = 999,500, Y(0) =500 )$ &2.015 (0.0038)& 2.015 (0.0035)  & 2.011 (0.0032) &2.000 (0.0028)  \\
$(X(0) = 9,950, Y(0) =50 )$& 2.032 (0.0221) & 2.032 (0.0222) &  2.025 (0.029) & 1.283 (0.2032) \\
$(X(0) = 990, Y(0)= 10 )$& 2.014 (0.1288)  & 2.014 (0.1289)   & 2.494 (0.3402) & 0.054 (1.9697) \\
$(X(0) = 99, Y(0)= 1 )$& -55 (1230)&  21 (41) & 1.471 (0.4831) & 2.168 (4.6)\\
\bottomrule
\end{tabular}%
}
\end{table}

\begin{figure}
\centering
  \centering
  \includegraphics[width = \textwidth]{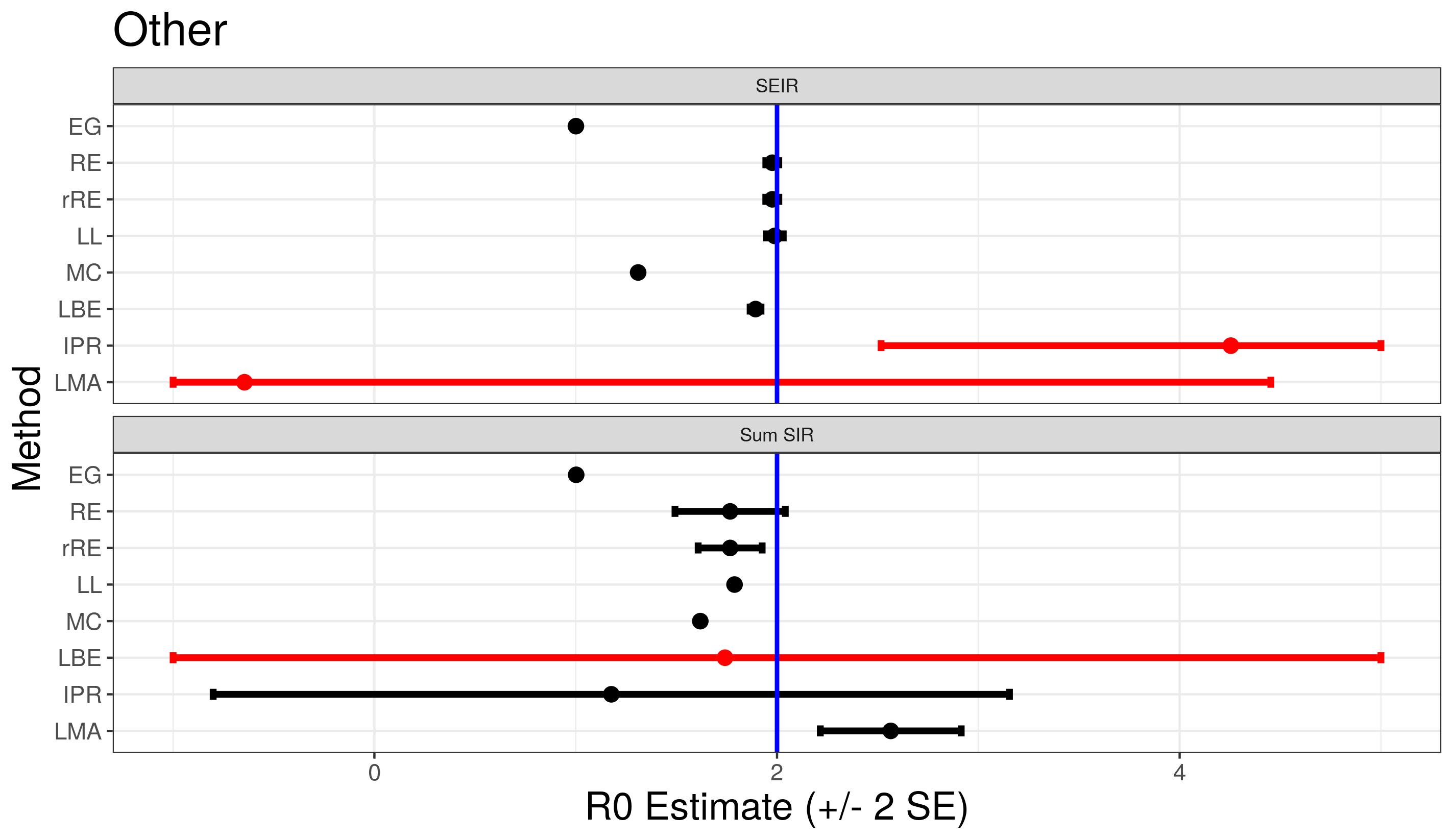}
\caption{Point and 95\% CI estimates of a result of a single simulated data set for the given parameter conditions.  The vertical blue line denotes the `true' value of $\rr$.  The estimates colored in red have CIs that are truncated at least on one side for visibility.  The corresponding table of estimates are shown in Table \ref{tab:non-sir-ests}.  The figure entitled ``SEIR'' refers to data generated according to Eq. \eqref{eq:seir}, and the data in ``Sum SIR'' refers to two SIR models that have been combined together, one with a true value of $\rr = 1.5$ and the other with $\rr = 2.5$. These forest plots correspond to the figures in Table \ref{fig:non-sir-ests}.}
    \label{fig:non-sir-ests}
\end{figure}

Finally, we show estimates and 95\% CIs where the simulated data is generated from non-Kermack and McKendrick SIR models in Figure \ref{fig:non-sir-ests} and Table \ref{tab:non-sir-ests}.  When we misidentify XEYZ (SEIR) data as XYZ (SIR) data, then we see we systematically (but very slightly) underestimate $\rr$.  While mathematically, we would expect underestimation of $\rr$, this is likely worse real life implications than overestimating $\rr$.  On the other hand, if there are two separate groups of equal size that have two different true values of $\rr$ (1.5 and 2.5), the combined estimate of the population $\rr$ is about 1.75.  This shows some of the non-linearity about the quantity $\rr$,  as it is not as simple as taking the average $\rr$ of the equally sized groups. This may indicate that correctly identifying sub-populations that have different risk factors is important to create accurate and informative estimates of $\rr$.  Overall, we find that proper model specification is very important when obtaining an estimate of $\rr$.

\begin{table}

\caption{\label{tab:non-sir-ests}Table of $\rr$ point estimates and standard error (SE) for non-Kermack and McKendrick SIR, simulated data. These numbers correspond to the forest plots in Figure \ref{tab:non-sir-ests}.}
\centering
\begin{tabular}{lll}
\toprule
\bf{Methods} & \bf{SEIR}\;$\widehat{\rr}$ $(SE(\widehat{\rr}))$  & \bf{SIR}$\times2$\; $\widehat{\rr}$ $(SE(\widehat{\rr}))$ \\
\midrule
EG & 1.001 ($<$1e-04) & 1.002 (2e-04)\\
RE & 1.977 (0.0161) & 1.767 (0.137)\\
rRE & 1.977 (0.0162) & 1.767 (0.0795)\\
LL & 1.989 (0.021) & 1.789 (0.0042)\\
MC & 1.31 (0) & 1.619 (0)\\
LBE & 1.893 (0.0135) & 1.742 (5.5501)\\
IPR & 4.254 (0.8687) & 1.177 (0.9889)\\
LMA & -0.645 (2.5491) & 2.565 (0.175)\\
\bottomrule
\end{tabular}
\end{table}

\subsection{2009 Pandemic Influenza Results}
\begin{figure}
	\centering
	\includegraphics[width=.8\textwidth]{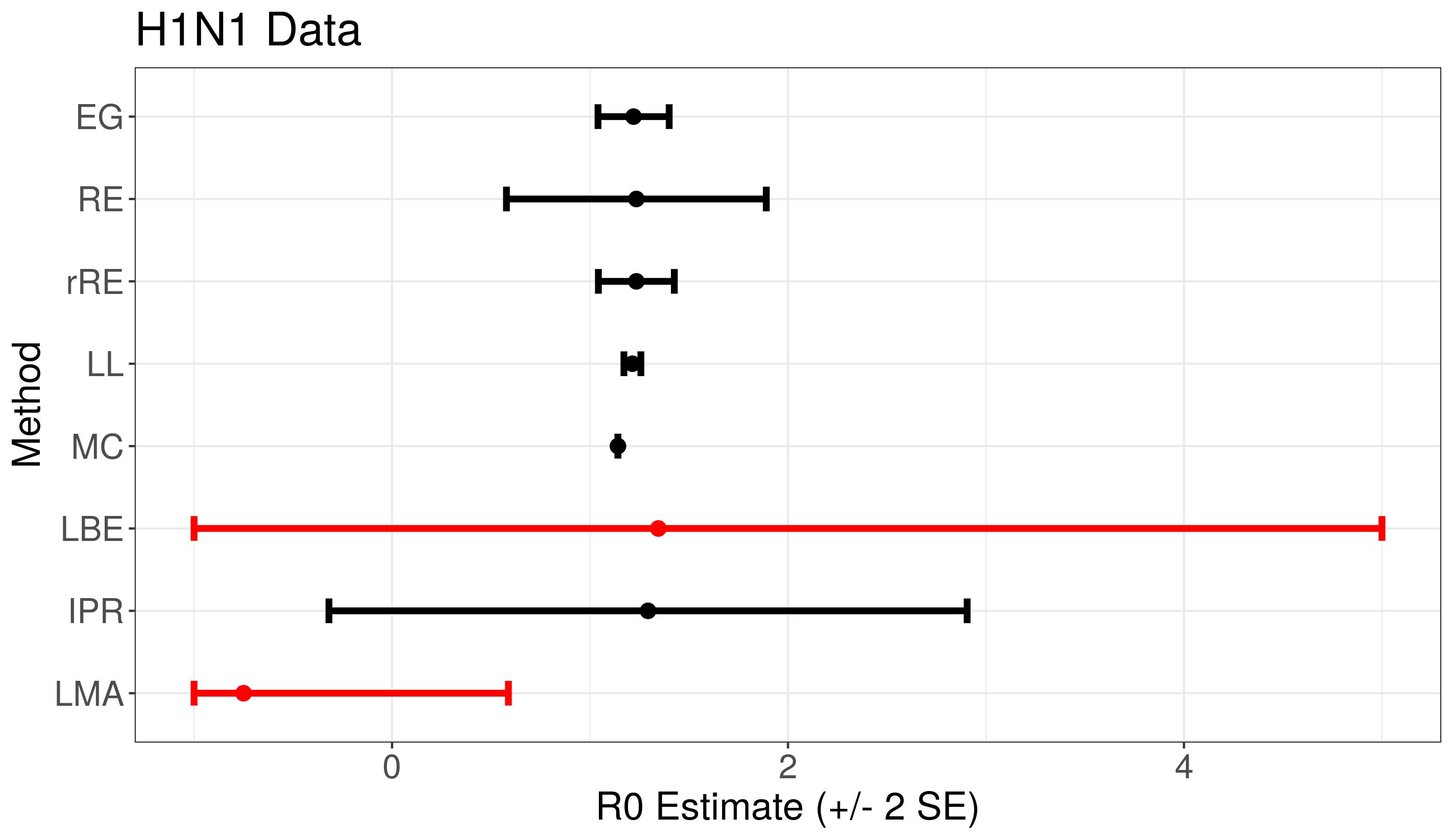}
	\caption{Results from SIR methods on H1N1 data, assuming $\gamma^{-1}$ = 3 days when imputing SIR data from the wILI reports.}\label{fig:h1n1-res}
\end{figure}

The forest plot for the \wxxsir method estimates (assuming $\gamma^{-1}= 3$ days when imputing SIR data from the wILI reports) is displayed in Figure \ref{fig:h1n1-res} and their corresponding table of values is seen in Table \ref{tab:h1n1-og}.  As this is real data, we cannot say what the true value of $\rr$ is, but encouragingly, the methods that had the best results in the simulation study (RE, rRE, LL, and LBE) have similar point estimates.  The median estimate is $\widehat{\rr} = 1.23$.  Surprisingly, even EG and MC have estimates close to 1.20, even though they rarely matched RE, rRE, LL, or LBE in our simulation study.  The full table of results is shown in Table \ref{tab:h1n1-og}.  Using the results from rRE,  we estimate that $\widehat{\rr} = 1.23$ with 95\% CI [1.03, 1.42], which is a smaller estimate than the value of 1.7-1.9 found in \cite{towers2009}, which used the same estimate of $\gamma^{-1} = 3$ days.

\begin{table}[H]
\caption{\label{tab:h1n1-og}Estimates of $\rr$ from the 8 methods and assuming $\gamma^{-1} = 3$ when imputing SIR data from the wILI reports.}
\centering

\begin{tabular}{@{}lrr@{}}
\toprule
\bf{Method} & \bf{$\widehat{\rr}$} & \bf{$SE(\widehat{\rr})$}\\
\midrule
EG & 1.220 & 0.0899\\
RE & 1.234 & 0.3281\\
rRE & 1.234 & 0.0959\\
LL & 1.214 & 0.0214\\
MC & 1.141 & 0.0000\\
LBE & 1.345 & 3.4319\\
IPR & 1.293 & 0.8059\\
LMA & -1.552 & 1.0699\\
\bottomrule
\end{tabular}%

\end{table}

\begin{figure}[h]
  \centering
  \includegraphics[scale = 0.45]{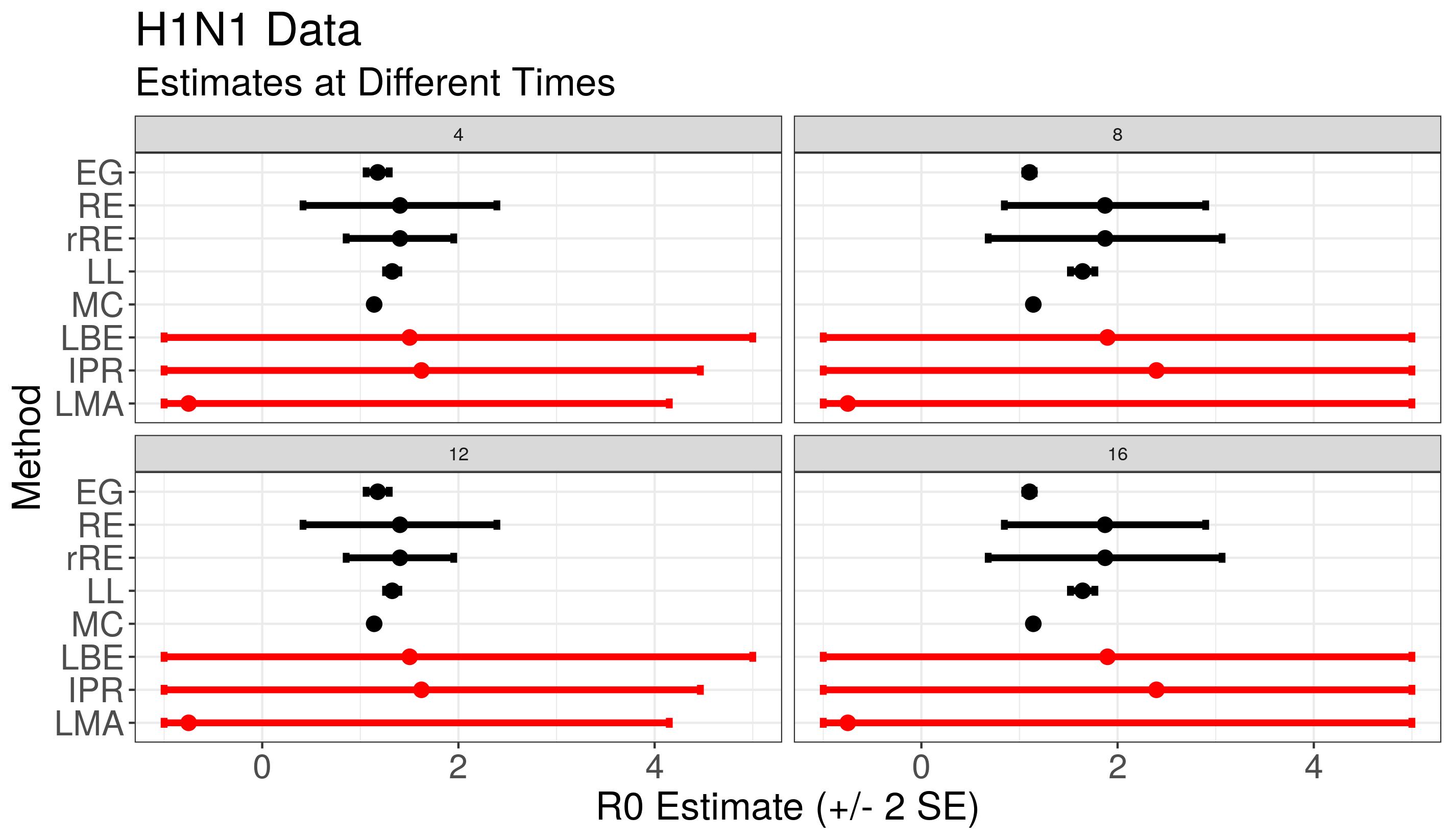}
  \caption{Results from SIR methods on H1N1 data, assuming $\gamma^{-1}$ = 3 days when imputing SIR data from the wILI reports.  This shows the estimates of $\rr$ using different maximum time points, to get a sense of how $\rr$ estimates can change with time.  The top left graph uses 4 weeks, the top right 8 weeks, bottom left 12 weeks, and bottom right 16 weeks worth of data.  The corresponding table of estimates is shown in Table \ref{tab:h1n1-time}.}\label{fig:h1n1-time}
\end{figure}

In Figure \ref{fig:h1n1-time}, we show the number of weeks used to estimate $\rr$, all with the initial time point of April 16, 2019 and using one of 4, 8, 12, or 16 weeks worth of data.  We see that the average estimate for most of the $\rr$ estimates are robust to the amount of time used but the estimates of $\rr$ from 4 weeks of data compared to the others, results in larger estimates of $\rr$.  The method estimates are reported in Table \ref{tab:h1n1-time}.  When using only 4 weeks of data, we see estimates of $\rr$ from LL to be 1.80 (0.33), which is much closer to the \cite{towers2009} estimate.

\begin{figure}[H]
  \centering
  \includegraphics[scale = 0.45]{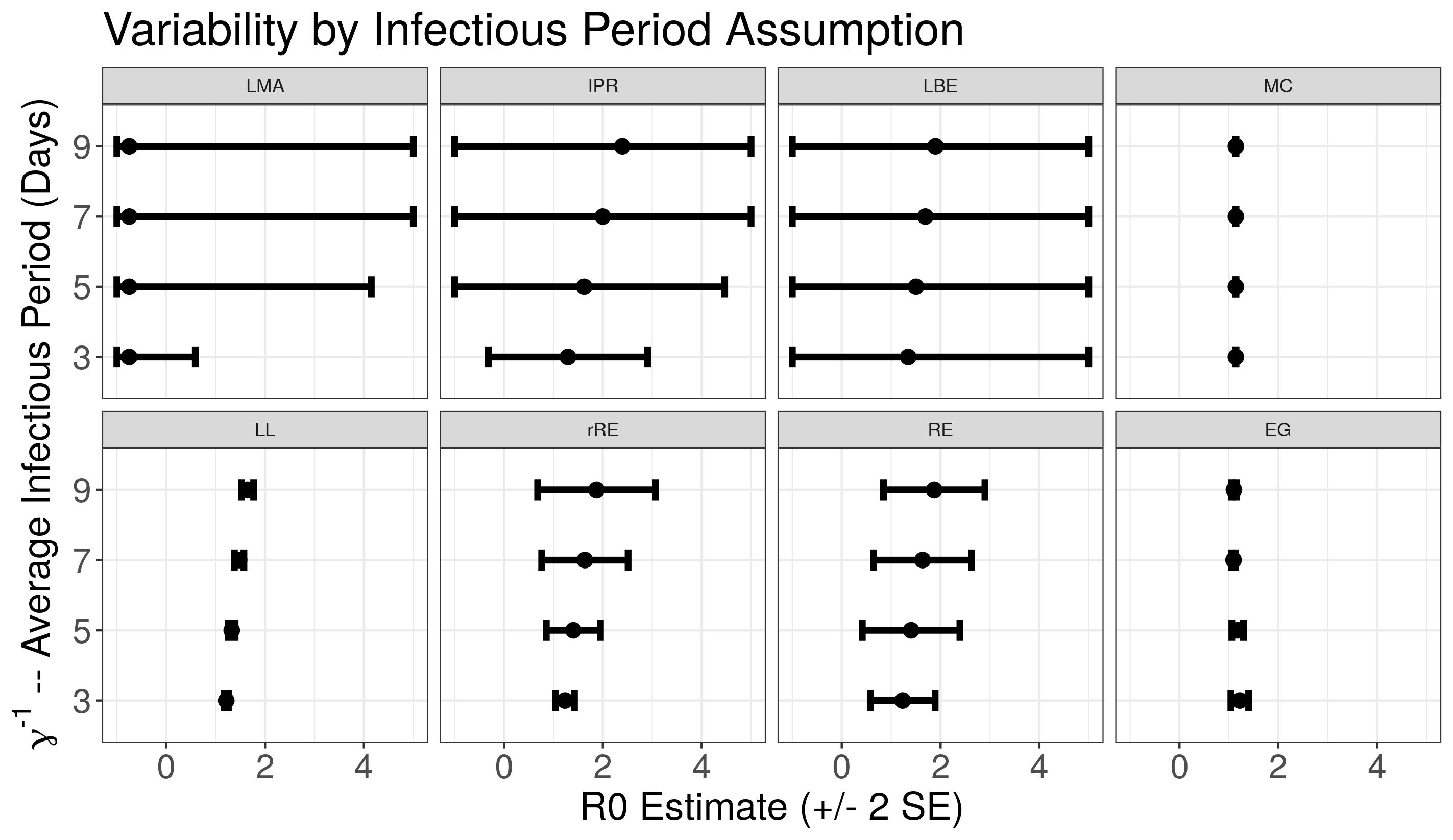}
  \caption{Variability of $\rr$ estimates for different values of $\gamma^{-1}$, the average time to recovery or average infectious period in imputing the SIR data from the wILI reports.}\label{fig:h1n1-gamma}
\end{figure}

\begin{table}[h]
\caption{Results from SIR methods on H1N1 data, assuming $\gamma^{-1}$ = 3 days when imputing SIR data from the wILI reports.  This shows the estimates of $\rr$ using different maximum time points, to get a sense of how $\rr$ estimates can change with time.  The table corresponds to the forest plots shown in Figure \ref{fig:h1n1-time}.}\label{tab:h1n1-time}
\centering
\begin{tabular}{@{}lllll@{}}
\toprule
Method & $\widehat{\rr}$--4 weeks &$\widehat{\rr}$--8 weeks & $\widehat{\rr}$--12 weeks & $\widehat{\rr}$--16 weeks\\
\midrule
EG & 1.336 (0.1678) & 1.167 (0.0592) & 1.117 (0.0339) & 1.098 (0.0245)\\
RE & 2.179 (0.2689) & 1.532 (0.2815) & 1.341 (0.3264) & 1.247 (0.3516)\\
rRE & 2.176 (0.5043) & 1.53 (0.2671) & 1.341 (0.1879) & 1.247 (0.1254)\\
LL & 1.796 (0.33) & 1.329 (0.0802) & 1.236 (0.0374) & 1.193 (0.0232)\\
MC & 1.141 ($<$1e-04) & 1.141 ($<$1e-04) & 1.141 ($<$1e-04) & 1.141 ($<$1e-04)\\
LBE & 1.885 (0.3735) & 1.427 (0.3735) & 1.317 (0.3735) & 1.276 (0.3735)\\
IPR & 2.376 (1.272) & 1.7 (1.1037) & 1.437 (0.9633) & 1.29 (0.8661)\\
LMA & -0.983 (8.1365) & 0.08 (4.5233) & 0.149 (3.4279) & -1.932 (6.9396)\\
\bottomrule
\end{tabular}
\end{table}
Our above results are generated after imputing the SIR data from wILI counts using the average time to recovery, $\gamma^{-1}$ = 3 days, which is a major assumption.  In Figure \ref{fig:h1n1-gamma}, it is clear that estimates of $\rr$ increase in magnitude as the estimate of $\gamma^{-1}$ increases, which makes sense as a longer infectious period leads to more infections.  Moreover, we see that the width of the CIs  typically increases for the estimates as $\gamma^{-1}$ increases and is especially noticeable for rRE.  On the other hand, we see that the LBE CI widths are large, which may indicate that the transmission spread does not follow the simple specified likelihood model well.  Finally, we see that both  EG and MC produce estimates that only increase by a few thousandths for different periods of infectiousness.  In contrast, the mean estimate from RE is 1.23 for $\gamma^{-1}= 3$ days and is 1.87 when $\gamma^{-1}$ = 9 days.  In context, the differences in estimates represent a final size (cumulative percent of infected individuals) difference from 35\% to 76\%, respectively, or approximately 123 million people in the US.

That said, all of RE, rRE, LBE, and LL have similar point estimates of $\rr$ for a given value of $\gamma^{-1}$ in the imputation step.  We also note that in all their estimates, RE and LBE have SE of at least 0.25, which means the confidence intervals are fairly large.  Of the methods we prefer, only LL has a reasonably small CI width for every value of $\gamma^{-1}$, with a maximum SE of 0.06.


Overall, from this analysis, we conclude when we assume that $\gamma^{-1}=3$ days, that $\rr$ is closer to 1.2 than it is to 1.6, which indicates that, at least within this time frame, $\rr$ for H1N1 is comparable to that of seasonal influenza where $\widehat{\rr}$ is estimated to be between 1.19-1.37 \citep{biggerstaff2014}. However, our estimate does not factor in how prevention/awareness campaigns influenced the spread of the disease, the differing seasonality of pandemic influenza (the time range we analyze is May-August) compared to seasonal influenza (the time range is typically October-March), and the severity and impact of the infection on the individual.  Another possible concern is that of ``backfill,'' the phenomenon of where the number of infections for the different weeks are constantly updated, even months after the original date.  As such, we have no guarantee we are using the same numbers from previous analyses.  Also, pandemic influenza occurred in two waves in the US, and we are analyzing the first wave, as was done in \cite{towers2009}. 

We assumed that $\gamma^{-1}$ = 3 days when imputing the SIR data from wILI reports.  If we had assumed a 7 or 9 day infectious period then our estimates of $\rr$ become more in line with those reported by \cite{biggerstaff2014} for pandemic influenza: estimates of $\widehat{\rr}$ between 1.3-1.7. Finally, we see that the estimates of $\rr$ are sensitive to the time period used (see Table \ref{tab:h1n1-time}) where 4 weeks of data lead to estimates of about 1.8 or larger and which substantially decreases as more data points (time) is added.

What this disparity in estimates indicates is how sensitive estimates of $\rr$ can be to different parameters and emphasizes the need for reproducible research in this area.  The reproduction number can be a viable way to compare disease outbreaks to one another, but it is not so simple as comparing one number to another, as the model framework (e.g. SIR vs. SEIR) must be considered.   We recommend, along with reporting the estimate and CI of $\rr$ to also explicitly report the modeling framework (e.g. Transmission Model: $S\to I \to R$, $\widehat{\rr} = 1.4,\; 95\%\; CI\; [1.1, 1.5])$ \textit{and} what steps are taken in pre-processing the data and what time steps are used and why.

\section{Discussion}\label{sec:discussion}

In this paper, we examine the nuances of the initial reproduction number $\rr$, with emphasis on 1) variability in methods of estimation, 2) sensitivity to a number of infection parameters, and 3) the role of data pre-processing steps to transform incidence counts to a more familiar data format.

As to issue 1), we examine \wxxsir methods to estimate $\rr$ which include exponential growth (EG), ratio estimator (RE), reparameterized ratio estimator (rRE), log linear (LL), Markov Chain (MC), likelihood-based estimate (LBE), incidence-to-prevalence ratio (IPR), and linear model approximation (LMA).  Although, this is not an exhaustive list of methods used to estimate $\rr$ from the SIR model, we believe it shows the diversity of the different methods that may be used to estimate this quantity.  We review the details of these methods in Sections \ref{sec:methods-eg}-\ref{sec:methods-lma}.




As to issue 2), we compare the the estimates of $\rr$ for the \wxxsir methods using a series of simulations.  To this effort, we generate different simulated data sets under different noise assumptions and estimate $\rr$ using each of these data sets.  We then analyze estimates of $\rr$ while varying individual parameters of the model including the infection and recovery parameters, the total amount of data, the initial percent of infectious and susceptible individuals, the standard deviation of the noise used to generate the simulated data, and the total number of individuals in a population.  We also analyze the model under misspecification, that is when a model other than the SIR model is used to generate the simulated data sets. Our simulation code is available for reproduction in our code repository: \url{github.com/skgallagher/r0}.

As to issue 3) we discuss one common approach to pre-processing incidence counts into the number of susceptible, number of infectious, and number of recovered individuals at each time step and how this pre-processing is dependent on our estimate of the average recovery period duration, $\gamma^{-1}$.  We show the effects of this pre-processing step in our application to 2009 pandemic influenza in the US.


Overall, we recommend based on our analysis:
\begin{itemize}
  \item RE, rRE, LL, and LBE provide the most accurate and informative estimates of $\rr$.  This is based on both the results of our simulation study and the similar results in the 2009 pandemic influenza data application.
  \item LL provides a way to establish a useful novel/secondary/confirmatory interpretation of $\rr$.
  \item LBE is a reliable and useful estimate and tends to generate larger CIs than the sum of square based methods (RE, rRE, LL).
  \item LBE is a simple likelihood based estimate and seems to work well in our simulation study and the data application, which shows to us that other more sophisticated LBE methods may work even better.
\item We do not recommend using EG or IPR even as `back of the envelope' calculations based on our simulation study.
  \item The larger the number of initially infectious individuals is, the smaller the CIs for $\rr$ are.
    \item Of the parameters described here, $\widehat{\rr}$ is most sensitive to changes in $T$, the time points used in an epidemic.
    \item Small sample sizes in total population size or the initial percent of infectious also cause disparity in $\rr$ method estimates and larger CIs than when changing other simulation parameters.

  \end{itemize}

  We also analyzed estimates of $\rr$ using data from the 2009 H1N1 pandemic in the USA.  From this study we conclude:
  \begin{itemize}
          \item For the H1N1 application, we find that when we assumed the infectious period was on average $\gamma^{-1}=3$ days, we found our estimates to be closer to $\widehat{\rr} = 1.2$, whereas the \cite{towers2009} reports estimates of 1.7-1.9 from.  In both our and their study, we use similar data imputation steps, most importantly assuming $\gamma^{-1}$=3 days.
    \item Also for H1N1, we found that when we increased $\gamma^{-1}$ to 7 or 9 days, we find estimates much more in line with 2009 pandemic influenza values of $\rr$ reported in \cite{biggerstaff2014}.
    \item From this, we conclude that imputation/pre-processing steps are essential in producing accurate and reliable estimates of $\rr$
    
  \end{itemize}

Overall, we find that $\rr$ is a difficult quantity to estimate.  We find that estimates often are nuanced and only interpretable in the context of other parameters, which is contrary to the idea of a one number summary, which $\rr$ is often purported to be.  We, however, are not yet ready to give up on the `most important quantity' in epidemic modeling.

To better understand what the process of estimating $\rr$ entails, we recommend the following guidelines.  First, we emphasize the need for reproducible research in this field, in every step there is: data collection, data cleaning, and statistical analysis.  We also recommend that every abstract/paper summary not only includes the estimate and 95\% CI of $\rr$ but also explicitly gives the disease framework and any pre-processing/imputation steps of the data.  This would go a long way for scientists to methodically scrape publication data to create an extensive table of $\rr$ for various outbreaks around the world.  Finally, we do not see any harm in presenting multiple estimates of $\rr$, especially those of rRE, RE, LL, and LBE.  Disparity in these estimates, for example, can show a deeper issue about the data.

The goal of this paper is explore some of the underlying nuances of $\rr$ how these nuances shape and guide ongoing statistical analysis.  We hope this paper provides a step in the right direction in a better understanding of $\rr$ for all scientists involved in the field of infectious disease transmission.

As this paper is being prepared, COVID-19, novel coronavirus, is beginning to severely outbreak in the US, and we see many estimates of $\rr$ being purported online and in the news (such as a table seen in \cite{midas2020}).  We hope this paper provides guidance in what features to analyze and assess when attempting to assess the reliability of $\rr$ estimates.


\bibliographystyle{apacite}      
\bibliography{references}   

\appendix

\section{Non Kermack and McKendrick SIR-Based   Data}\label{app:sim-data}

\textbf{XYZ (SIR) $\times$ 2}
In this model, we have two independent XYZ models that are then aggregated together.  Specifically, both models are given from the equations in Eq. \eqref{eq:sir}.  The first one, denoted $A$, is given by $(X_A(0) = 49,950, Y_A(0) = 50, Z_A(0) = 0)$ and disease parameters such that $\rr^{(A)} = 1.5$ ($\beta_A = 0.015, \gamma_A = 0.01)$.  The  second one is denoted by $B$ and is given by $(X_B(0) = 49,950, Y_B(0) = 50, Z_B(0) = 0)$ and disease parameters such that $\rr^{(B)} = 2$ ($\beta_A = 0.02, \gamma_A = 0.01)$.  We then add the two sets of compartments together: $$(X(t) = X_A(t) + X_B(t), Y(t) = Y_A(t) + Y_B(t), Z(t) = Z_A(t) + Z_B(t)).$$  Then following the equations in Eq. \eqref{eq:sim-models}  $f= X$ and $g = Z$.\\

\noindent \textbf{XEYZ (SEIR)}.  The deterministic SEIR (XEYZ) model (seen in \cite{cintronarias2009}) is based upon the following ODEs:

\begin{align}\label{eq:seir}
  \frac{dX}{dt} &= - \frac{\beta XY}{N} \nonumber \\
  \frac{dE}{dt} &= \frac{\beta XY}{N}  - \mu E\nonumber \\
  \frac{dY}{dt} &= \mu E - \gamma Y \nonumber \\
  \frac{dZ}{dt} &= \gamma Y,
\end{align}
again, with a constant population of $N$ and with known initial values $(X(0), E(0), Y(0), Z(t))$.  As in the XYZ model, $\beta$ and $\gamma$ represent the average rate of infection and average rate of recovery, respectively.  The parameter $\mu$ is the rate at which latently exposed individuals become infectious.   The reproduction number for the XEYZ model, is $\rr = \frac{\beta}{\gamma}$, like in the XYZ model.

To simulate XYZ data from the XEYZ model, we first determine the XEYZ values for a given set of initial values.  Then we combine the susceptible and exposed states into a new susceptible state, $X^{new}(t) = X(t) + E(t)$.  Then following the equations in Eq. \eqref{eq:sim-models}  $f= X^{new}$ and $g = Z$.  In the simulations we let $\mu = 0.1$.

\end{document}